\mathsurround 2pt
\baselineskip 18pt plus 2pt

\def\zero{\footline={\ifnum\pageno>0
\hss\folio\hss\fi}
\pageno=0}

\def\bx{\hfill{\vbox{\hrule\hbox{\vrule\kern6pt\vbox{\kern6pt}\vrule}\hrule}}}

\def\ref #1#2{\noindent\hangindent=4em\hangafter1
                \hbox to4em{\hfil#1\quad}#2}

\parskip=6pt
\zero

\centerline {\bf COULOMB FIELD OF AN ACCELERATED CHARGE:}

\centerline {\bf PHYSICAL AND MATHEMATICAL ASPECTS}
\bigskip

\bigskip

\centerline {Francis J. Alexander}
\centerline {Department of Physics and Astronomy}
\centerline {Rutgers University}
\centerline {Piscataway, NJ  08855}
\bigskip

\bigskip

\centerline {and}
\bigskip

\bigskip

\centerline {Ulrich H. Gerlach}
\centerline {Department of Mathematics}
\centerline {Ohio State University}
\centerline {Columbus, Ohio  43210}
\centerline {gerlach@math.ohio-state.edu}
\vfill\eject

\centerline {\bf ABSTRACT}
\bigskip

The Coulomb field of a charge static in an accelerated frame has
properties that suggest features of electromagnetism which are
different from those in an inertial frame.

An illustrative calculation shows that the Larmor radiation reaction
equals the electrostatic attraction between the accelerated charge and
the charge induced on the surface whose history is the event horizon.

A spectral decomposition of the Coulomb potential in the accelerated
frame suggests the possibility that the distortive effects of this
charge on the Rindler vacuum are akin to those of a charge on a
crystal lattice.  The necessary Maxwell field equations relative to
the accelerated frame, and the variational principle from which they
are obtained, are formulated in terms of the technique of geometrical
gauge invariant potentials. They refer to the transverse magnetic (TM)
and the transeverse electric (TE) modes.

\vfill\eject

\centerline {\bf I. MOTIVATION AND INTRODUCTION.}

The classical and quantum mechanical pictures of a charged particle together with its Coulomb field
are well known in an inertial frame.

The classical picture is the one where the static charge is the source of its spherical 
electric field which can be derived from the electrostatic potential.  
The quantum mechanical picture is the one where the charge is surrounded by a cloud of 
virtual quanta$^{1, 2}$ each one of which is emitted and reabsorbed by the charge.  The
(self)interaction energy due to these processes gives rise to the ``renormalized'' experimentally
observed restmass energy of the charged particle.  If a second charged particle is present then there
are two clouds of (virtual) quanta.  In this case there is a probability that a quantum emitted
by one charge can be absorbed by the other, and vice versa.  The mutual interaction energy due to such
exchange processes is given by  
$$\eqalignno{V(\vec X_1, \vec X_2) &= e_1e_2{e^{-m\vert\vec X_1-\vec x_2\vert}\over 
{\vert \vec X_1 -\vec X_2\vert}}&(1.1)\cr
&=4\pi e_1e_2\int\!\!\int\!\!\int {1\over k^2+m^2}e^{i\vec k\cdot(\vec X_1-\vec
X_2)}{d^3k\over (2\pi)^3}\cr}$$

\noindent the familiar scalar potential, which for $m=0$ reduces to a Coulomb field.  
A key ingredient
to this  result is that the quanta responsible for this interaction are the familiar Minkowski
quanta, the elementary modifications (= ``excitations'') of the familiar translation invariant
Minkowski vacuum.  

The question now is this:  Does the quantum mechanical picture of the exchange interaction between a
pair of charges extend to an accelerated frame?  In other words, can the classical potential
energy between two uniformly accelerated charges (in the same accelerated frame) still be attributed
to the exchange of virtual quanta in the accelerated frame?

To answer this question it is not enough to restrict one's attention to the 
quantum mechanics based on the Minkowski vacuum and its excitations (= mesons ``photons'').  The
ground state of an acceleration-partitioned field is entirely different from the Minkowski vacuum. 
Indeed, that ground state determines a set of quantum states which is disjoint from
(i.e. unitarily unrelated to) the set of quantum states based on the Minkowski vacuum state$^3$. 
Physically that ground state has the attributes which are reminiscent of a condensed vacuum
state$^4$.

Very little is known about interactions between particles and an acceleration-partitioned 
field in its (condensed) ground state.  The
contrast between such interactions and those that are based on the Minkowski vacuum state of the
field raises some non-trivial issues of principle which are not answered in this paper however.

Here we shall erect the framework that allows a very economic analysis of the interaction between
currents and fields.  The system is, of course, the classical Maxwell field with its charged
sources.  The utility of this framework lies in the fact that the four-dimensional problem has been
reduced to two dimensions in such a way that the field and the charge current can be viewed relative
to any linearly accelerated coordinate frame.

This paper accomplishes four tasks.

\noindent 1.  Sections II -- III formulate the full classical Maxwell electrodynamics in 
terms which are most natural for a linearly accelerated coordinate frame.  This is done by exhibiting
a reduced (``$2+2$'') variational principle and the concomitant reduced set of decoupled inhomogeneous
wave equations for the to-be-quantized transverse-electric (TE) and transverse-magnetic (TM) degrees
of freedom.  The reduction procedure is not new.  It has already been applied to the linearized
Einstein field equation of a spherically symmetric spacetime$^5$.

\noindent 2.  Sections IV -- V give what in an inertial frame corresponds to a multipole expansion.  
Section VI reviews the well known quantum mechanical picture of the interaction between two charges
as a spectral sum of exchange processes, and then gives a spectral decomposition of the Coulomb
potential between a pair of linearly uniformly
accelerated charges.  

\noindent 3.  Sections VII suggests that the Coulomb attractive force between an accelerated charge
and the induced charge on its event horizon be identified with Larmor's radiation reaction force.

\noindent 4.  Section VIII compares a pair of charges static in the vacuum of an accelerated frame
to two polarons, and then draws attention to the possibility that their interaction might be
different from what one expects from quantum mechanics relative to the inertial vacuum.

\vfill\eject

\centerline {\bf II.  THE REDUCED VARIATIONAL PRINCIPLE.}
\bigskip

Classical Maxwell electrodynamics is a consequence of the principle of extremizing the action 
integral
$$\eqalignno {I~[A_\mu ] &=\int \int \int \int \lbrace {-1 \over {16\pi}} (A_{\nu;\mu} - A_{\mu ;\nu})
(A^{\nu ;\mu} - A^{\mu ;\nu}) + J^\mu A_\mu \rbrace \sqrt {-g}~d^4x.&(2.1)\cr}$$ 

\noindent The resulting Euler equations is the familiar system of inhomogeneous Maxwell wave 
equations
$$\eqalignno {A^{\mu ;\nu}~_{;\mu} - A^{\nu ;\mu}~_{;\mu}& = 4\pi J^\nu.&(2.1)\cr}$$

\noindent These equations imply charge conservation,
$$\eqalignno {J^\nu~_{;\nu} &=0. &(2.3)\cr}$$

\noindent But this fact also follows directly from the demand that $I$ be invariant under 
gauge transformations $A_\mu \to A_\mu + \Lambda_{,\mu}$ i.e. from
$$\eqalignno {I~[A_\mu + \Lambda_{,\mu}]&= I~[A_\mu ].&(2.4)\cr}$$
\bigskip

\centerline {\bf A. Scalar and Vector Harmonics.}

The fact that $y-z$ plane accomodates the Euclidean group of symmetry operations
implies that one can introduce various sets of scalar and vector harmonics which have simple
transformation properties under the various group operations.  We shall use the complete set of delta
function normalized scalar harmonics $$\eqalignno {Y^k~(y,z) &= {1 \over {2\pi}}
e^{i(k_yy+k_zz)}&(2.5)\cr}$$

\noindent and the corresponding two sets of vector harmonics,
$$\eqalignno {Y^k_{,a} \equiv {\partial Y^k \over {\partial x^a}} &= ik_aY^k&(2.6a)\cr
\noalign{\hbox{and}}
Y^k_{,c} \epsilon^c~_a &= ik_c\epsilon^c~_aY^k.&(2.6b)\cr}$$

\noindent Here $k \equiv (k_y, k_z)$ identifies the harmonic.

The ``$a$'' refers to the coordinates $x^2 = y$ or $x^3 = z$, which span $R^2$, the transverse 
symmetry plane on which the Euclidean group acts.  The expression $\epsilon^c\,_a$ is the
antisymmetric Levi-Civita symbol.

Instead of this set of translation eigenfunctions one could just as well have used 
the complete set of scalar Bessel harmonics (rotation eigenfunctions), 
$$Y^{km}(r,\theta )= J_m(kr)e^{im\theta},~~~~ k = \sqrt {k^2_y + k^2_z}$$

\noindent and the concomitant set of vector harmonics.  Indeed, one can equally well 
use an orthonormalized {\it discrete} set of trigonometric or Bessel harmonics on a finite rectangle
or a disc in the $y-z$ plane.  Which of all these possibilities one must choose depends, of course,
entirely on the boundary conditions which the Maxwell field satisfies in the $y-z$ plane.  However
the reduced form of the variational principle and the inhomogeneous Maxwell wave equations (See Eqs.
(3.1) for the $TM$ modes and (3.2) for the $TE$ modes) have the same form for all these different
harmonics.

For the sake of concreteness we shall use the familiar translation scalar and vector 
harmonics, Eq. (2.5) and (2.6).  All the normalization integrals for the vector harmonics follow
directly from 
$$\eqalignno {\int \int (Y^k)^{\ast} Y^{k'}dydz &= \delta (k_y - k'_y)\delta (k_z - k'_z) \equiv
\delta^2 (k-k').&(2.7)\cr}$$ 

\noindent Thus one has, for example,
$$\eqalignno {\int \int (Y^k)^{\ast}_{,a}Y^{k'}_{,b}dydz &=k_a k_b \delta^2 (k-k')&(2.8a)\cr
\int \int (Y^k)^{\ast}_{,c} \epsilon^c~_b Y^{k'}_{,d} \epsilon^{db} dy dz&= k^2 \delta^2 (k-k');
~~~~ k^2 = k^2_y + k^2_z&(2.8b)\cr}$$

\noindent while the ``longitudinal'' and the ``transverse'' vector harmonic are always orthogonal, 
$$\eqalignno {\int \int (Y^k)^{\ast}_{,a} Y^{k'}_{,b} \epsilon^{ba} dy dz &= 0.&(2.9)\cr}$$

\noindent All integrations are over $R^2$, the $y-z$ plane.
\bigskip

\centerline {\bf B.  Transverse Manifold $R^2$ and Longitudinal Manifold $M^2$}

In order to reduce the variational principle and the Maxwell field equations, one expands 
the vector potential and the current density in terms of these scalar and vector harmonics.  The
$4$-vector potential is  
$$\eqalignno {A_\mu(x^\nu )&= (\,A_B(x^\nu ), A_b(x^\nu )\,)&(2.10)\cr
&= (\,\mathop{\textstyle{\sum}}\limits_k a^k_B(x^C)Y^k,~~\mathop{\textstyle{\sum}}\limits_k a^k(x^C){\partial Y^k \over {\partial
x^b}} + A^k(x^C)  {\partial Y^k \over {\partial x^d}}\epsilon^d~_b\,)\cr}$$

\noindent and the $4$-current density is
$$\eqalignno {J_\mu (x^\nu )&= (~J_B(x^\nu ), J_b(x^\nu )~)&(2.11)\cr
&= (~\mathop{\textstyle{\sum}}\limits_k j^k_B(x^C)Y^k,~~~\mathop{\textstyle{\sum}}\limits_k~j^k(x^C){\partial Y^k \over {\partial x^b}} + J^k(x^C)
{\partial Y^k \over {\partial x^e}} \epsilon ^e~_b\,)\cr}$$

\noindent Here
$$\eqalignno {\mathop{\textstyle{\sum}}\limits_k\,(\cdots ) &\equiv \int^\infty_{-\infty} \int^\infty_{-\infty} {d^2k 
\over {2\pi}}(\cdots)&(2.12)\cr}$$

\noindent is the mode integral over the harmonics.  The scalar (on $R^2)$ harmonic expansion 
coefficients are $a^k_B (x^C)$ and $j^k_B(x^C)$.  They are components of {\it vectors} on $M^2$, 
the $2$-dimensional Lorentz spacetime spanned by $x^C = (x^0, x^1)$.  Similarly the expansion 
coefficients for the vector (on $R^2)$ harmonics $Y^k_{,b}$ are $a^k(x^C)$ and $j^k(x^C)$, while 
those for $Y_{,c}\epsilon^c~_b$ are $A^k(x^C)$ and $J^k(x^C)$.  All these coefficients are 
{\it scalars} on $M^2$.  Evidently the $4$-dimensional Minkowski spacetime $M^4$, which is 
coordinatized by $x^\nu = (x^0 , x^1, x^2, x^3)$, has been factored by the symmetries transverse 
to the acceleration into the product 
$$M^4 = M^2 \times R^2.$$

\noindent Here $M^2$ is coordinatized by $x^C = (x^0, x^1)$ and $R^2$ by $x^b = (x^2, x^3) = 
(y,z)$.  The metric of $M^4$ relative to the accelerated coordinate frame has the form
$$\eqalignno {ds^2&= g_{\mu \nu}dx^\mu dx^\nu&(2.13)\cr
&= g_{AB}(x^C)dx^Adx^B+g_{ab}(x^d)dx^a dx^b\cr}$$

\noindent It is block diagonal
$$[g_{\mu \nu}] = \left[ \matrix {g_{AB}&0\cr
                                  0&g_{ab}\cr}\right] .$$

Consequently $M^2$ and $R^2$ are mutually orthogonal submanifolds.  We shall call $R^2$ 
the {\it transverse} submanifold because it is perpendicular to the world line of a linearly
accelerated charge.  The geometric objects intrinsic to it are its metric tensor field, 
$$\eqalignno
{g_{ab} dx^a dx^b &= dy^2 + dz^2,&(2.14)\cr}$$

\noindent and the scalar and vector fields given by
$$Y^k(x^a),~~ {\partial Y^k \over {\partial x^b}},~~ {\rm and }~
{\partial Y^k \over {\partial x^c}} \epsilon^c\,_b\,.$$

We shall call $M^2$ the {\it longitudinal} submanifold because it contains the world 
line of a linearly accelerated charge.  The geometric objects intrinsic to it are not only its metric
tensor field $$\eqalignno {g_{AB}& (x^C)dx^Adx^B&(2.15)\cr}$$

\noindent but also the scalar and vector fields given by the coefficients of the harmonics in Eqs.
(2.10) and  (2.11).  Relative to a linearly and uniformly accelerated coordinate frame given by
$$\eqalignno {t &= \xi \sinh\,\tau &(2.16)\cr
x&= \xi \cosh \tau\cr}$$

\noindent the metric of the longitudinal submanifold $M^2$ has the form
$$\eqalignno {g_{AB} dx^A dx^B &= - \xi^2 d\tau^2 + d\xi^2.&(2.17)\cr}$$

In general however, the coordinate frame is not uniformly accelerating, and the 
metric does not have a correspondingly simply form. 

Our task of ``reducing'' the Maxwell field equations and its variational principle 
consists of formulating them strictly in terms of geometrical objects defined solely on $M^2$. 
Roughly speaking, we ``factor out'' the $y-z$ dependence of each harmonic degrees of freedom.  Thus we
introduce the harmonic expansions Eqs. (2.10) and (2.11) into the Maxwell wave Eqs. (2.2) and equate
the coefficients of the corresponding scalar and vector harmonics.  The result is the reduced set of
Maxwell wave Eqs. (3.1) and (3.2) on $M^2$.

The reduction of the variational principle is more informative because it directly 
relates gauge invariance to the structure of the wave equations.  Thus introduce Eqs. (2.10) and
(2.11) into the action integral, Eq. (2.1).  Using the fact that  $$\eqalignno {\sqrt {-g}d^4x &=
\sqrt {-g^{(2)}} d^2xdydz&(2.18)\cr \noalign{\hbox{with}}
g^{(2)}&= det [g_{AB}],\cr}$$

\noindent do the integration over $R^2$,
$$\int^\infty_{-\infty}\int^\infty_{-\infty}(\cdots )dy\,dz.$$

Finally make use of the orthogonality and the normalization integrals Eqs. (2.7) - (2.9).  
After a straightforward evaluation, the action integral decomposes into two independent sums over
each of the familiar transverse magnetic ($TM$, no magnetic field along the $x$-direction) and the
transverse electric ($TE$, no electric field along the $x$-direction) modes, $$\eqalignno {I~ &=
\sum_k I^k_{TM} + \sum_k~ I^k_{TE}.&(2.19)\cr}$$ 

\noindent Here the mode ``summation'' is given by Eq. (2.12).  The action for a $TM$ 
mode of type $k\equiv (k_y, k_z)$
$$\eqalignno {I^k_{TM}&= {1 \over {4\pi}} \int \int \lbrace - {1 \over {4}} (
{\cal {A}}_{B,C} - {\cal {A}}_{C,B})({\cal {A}}^\ast_{D,E} - {\cal {A}}^\ast_{E,D})
~~g^{BD}g^{CE}&(2.20)\cr
\cr
&- {1 \over {2}} k^2 {\cal {A}}^B{\cal {A}}^\ast_B + 4\pi j^Ba^\ast_B + 4\pi k^2ja^\ast 
\rbrace~~ \sqrt {-g^{(2)}}~d^2x,\cr}$$

\noindent where
$$\eqalignno {{\cal {A}}_B&= a^k_B - a^k_{,B}&(2.21)\cr
\noalign{\hbox{and}}
{\cal {A}}^\ast_B&= a^{-k}_B - a^{-k}_{,B}\cr}$$

\noindent refers to the mode $-k\equiv (-k_y, -k_z)$.

Because the total Maxwell field is real, $k \to -k$ corresponds to taking the complex 
conjugate of an amplitude.  Thus denoting $Y^\ast_B$ as the complex conjugate of $Y_B$ is consistent.

The action for a $TE$ mode by contrast is
$$\eqalignno {I^k_{EM} &= {k^2 \over {4\pi}}\int \int \lbrace - {1 \over {2}} 
A_{,B}~a^\ast_{,C}~~g^{BC} - {1 \over {2}} k^2 AA^\ast + 4\pi JA^\ast \rbrace \sqrt {-g^{(2)}}d^2x .
&(2.22)\cr}$$

\noindent Here we suppressed superscripts by letting 
$$\eqalign {A &= A^k,\cr
\noalign{\hbox{and we set}}
A^\ast &= A^{-k}\cr}$$

\noindent because the total Maxwell field is real.
\vfill\eject

\centerline {\bf III.  REDUCED MAXWELL WAVE EQUATION.}

It is now straightforward to obtain the reduced Maxwell field equation by extremizing 
the action.  For the $TM$ modes one has$^5$
$$\eqalign {0 ={\delta I \over {\delta a^\ast_B}}&:\cr&\cr 
0= {\delta I \over {\delta a^\ast}}&:\cr}
\eqalign {({\cal {A}}_{C\mid B}- {\cal {A}}_{B\mid C})^{\mid C} + k^2 {\cal {A}}_B & = 4\pi j_B\cr&\cr
{\cal {A}}^B_{\mid B}& = 4\pi j\cr}$$
\vskip -.9in \hfill $(3.1a)$
\vskip .1in \hfill $(3.1b)$
\bigskip

\noindent For the $TE$ modes one has
$$\eqalignno {0 = {\delta I \over {\delta A^\ast}}&:\qquad \qquad - A_{,B}~^{\mid B} + k^2 A = 4  
\pi J&(3.2)\cr}$$

\noindent Here the vertical bar ``$\mid$'' means covariant derivative obtained 
from the metric 
$$g_{AB}dx^Adx^B$$ 
on $M_2$.  It is clear that Eqs (3.1) and (3.2) are
geometrical vector and scalar equations on $M_2$.  They are equivalent to the unreduced Maxwell wave
Eqs. (2.2).  The reduced charge conservation equation corresponding to Eq. (2.3) is $$\eqalignno
{j^B_{\mid B} - k^2j = 0&&(3.3)\cr}$$

\noindent It is obtained from the divergence of Eq. (3.1a) combined with Eq. (3.1b).

\noindent {\bf A.  Gauge Invariance and Charge Conservation.}

The requirement that the action $I$ be invariant under the gauge transformation
$$A_\mu \to \overline {A}_\mu = A_\mu + \Lambda_{ , \mu}$$

\noindent gives rise to charge conservation, Eq. (2.3).  The gauge scalar $\Phi $ has the form
$$\Phi = \Sigma~\phi^k(x^C)\,Y^k$$

\noindent One sees from Eq. (2.10) that it induces the following changes on the vectors and 
scalars on $M^2$ for each mode (we are suppressing the mode index $k$)
$$\eqalign {a_B \to \overline {a}_B& = a_b + \phi_{,B}\cr
a\to \overline {a}& =a + \phi\cr
{\cal {A}}_B \to \overline {\cal {A}}_B &= \overline {a}_B - \overline {a}_{,B} = {\cal {A}}_B\cr
A \to \overline {A} &= A\cr}$$

Thus one sees that ${\cal {A}}_B$ and $A$ are gauge {\it invariant } geometrical objects 
on $M^2$, while $a_B$ and $a$ are gauge {\it dependent} objects on $M^2$.  If one demands that the
reduced action, Eqs (2.19), be gauge invariant, i.e.  $$I~[a_B+\phi_{,B}; a + \phi ; A] = I~[a_B; a;
A],$$

\noindent then one has 
$$j^B~_{\mid B} - k^2j = 0,$$

\noindent i.e. charge conservation for each mode.  It is clear that if the action is required 
to be an extremum under arbitrary variations, i.e.
$$I~[a_B + \delta a_B; a+\delta a; A + \delta A] = I~[a_B; a; A],$$

\noindent and thereby gives rise to the Maxwell field Eqs. (3.1) and (3.2) then charge 
conservation is guaranteed.  This is so because a gauge transformation is merely a special type of
variation which keeps $I$ stationary.

What is not so clear at first is why the Maxwell wave Eqs (3.1) and (3.2) are manifestly 
gauge invariant, while the action $I_{TM}$ in Eq. (2.20) does not enjoy this property.  The
offending terms in that integral are $$\int \int (j^Ba^\ast_B+k^2j~a^\ast )~~\sqrt
{-g^{(2)}}d^2x.$$

\noindent If one {\it assumes} charge conservation, i.e. Eq. (3.3) then these offending 
terms becomes
$$\eqalign {&=\int \int (j^B a^\ast_B + j^B~_{\mid B} a^\ast )~~ \sqrt {-g^{(2)}} d^2x\cr
&= \int \int j^B(a^\ast_B - a^\ast_{,B})~~ \sqrt {-g^{(2)}}d^2x = \int \int j^B
{\cal {A}}^\ast_B~~\sqrt {-g^{(2)}}d^2x\cr}$$

\noindent which is manifestly gauge invariant.  We therefore see that the {\it manifestly} 
gauge invariant action functional
$$\eqalignno {I^k_{TM} + I^k_{TE}&= {1\over 4\pi}\int \int \lbrace - {1 \over {2}}({\cal {A}}_{B\mid
C} -  {\cal {A}}_{C\mid B}) {\cal {A}}^{\ast B\mid C}- {1 \over {2}}k^2 {\cal {A}}^B
{\cal {A}}^\ast_B 
+ 4\pi j^B{\cal {A}}^\ast_B\rbrace \sqrt {-g^{(2)}}d^2x~~~~&(3.4)\cr
&+{k^2 \over {4\pi}}\int \int \lbrace - {1\over {2}}A_{,B}A^\ast_{,C}~~ g^{BC} - 
{1 \over {2}} k^2 AA^\ast + 4\pi JA^\ast \rbrace \sqrt {-g^{(2)}}d^2x\cr}$$

\noindent yields all the Eqs. (3.1) and (3.2) except one, namely Eq. (3.1b).  To obtain it, 
charge conservation has to be assumed explicitly; it can not be obtained from the manifestly
gauge invariant action functional.

\noindent {\bf B. The Electromagnetic Field}

The $TE$ field modes and the $TM$ field modes are totally decoupled from each other 
and thus evolve independently.  It is easy to obtain the electromagnetic field.  It decomposes
into blocks $$\eqalignno {\lbrack F_{\mu\nu} \rbrack &\equiv \lbrack A_{\nu , \mu } - A_{\mu ,
\nu}\rbrack = \left[ \matrix {F_{BC}&F_{Bb}\cr
F_{bB}&F_{bc}\cr}\right] .&(3.5)\cr}$$

\noindent With the help of Eq. (2.10) the Maxwell field of a typical $TE$ mode has the form
$$\eqalignno {\lbrack F^k_{\mu \nu}\rbrack_{TE} &=\left[ \matrix {0&~A_{,B}Y^k_{,d}~~
\epsilon^d~_b\cr
-A_{,B}Y^k_{,d}~~\epsilon^d~_b&-Ak^2~~Y^k\epsilon_{bc}\cr}\right].&(3.6)\cr}$$                                 
\noindent Here the gauge invariant scalar $A$ satisfies the inhomogeneous $TE$ wave Eq. (3.2),
$$-A^{\mid C}~_{,C}+k^2A = 4\pi J.$$

\noindent The Maxwell Field of a typical $TM$ mode has with the help of Eq. (2.10) the form
$$\eqalignno {\lbrack F^k_{\mu \nu}\rbrack_{TM} &= \left[ 
\matrix {({\cal {A}}_{C,B} - {\cal {A}}_{B,C})Y^k&-{\cal {A}}_B Y^k_{,b}\cr
{\cal {A}}_B Y^k_{,b}&0\cr}\right]&(3.7)\cr}$$

\noindent The gauge invariant vector potential ${\cal {A}}_B$ on $M^2$ can readily be 
obtained by decoupling the inhomogeneous wave Eqs. (3.1).  Observe that 
$$\eqalignno {{\cal {A}}_{C\mid B} - {\cal {A}}_{B\mid C} &= \Phi \epsilon_{CB}&(3.8)\cr
\noalign{\hbox{where}}
\Phi&= -{\cal {A}}_{E,F}\epsilon^{EF}&(3.9)\cr}$$

\noindent This quantity is a scalar on $M^2$ and it is the longitudinal electric field 
amplitude of a $TM$ mode.  It is not to be confused with the gauge scalar in the previous subsection
A.  The quantity $\epsilon_{CB}$ are the components of the totally
antisymmetric tensor on $M^2$.  Multiply both sides of Eq. (3.1a) by $\epsilon^{BD}$, use
$$\eqalignno {\epsilon_{CB}\epsilon^{BD}&= \delta^D_C&(3.10)\cr}$$

\noindent and take the divergence of both sides of Eq. (3.1a).  The result is the master $TM$
wave equation on $M^2$: 
$$\eqalignno {-\Phi_{,D}~^{\mid D}+k^2\Phi &= 4\pi j_{B\mid
D}\epsilon^{BD}&(3.11)\cr}$$

\noindent From its solution one can recover all 
components of the $TM$ electromagnetic field in Eq. (3.7).  Indeed, the gauge invariant vector
potential ${\cal {A}}_B$ is obtained from the vector Eq. (3.1a).  Combining it with Eq. (3.8) one
obtains $${\cal {A}}_B= \lbrack 4\pi j_B + \Phi_{,C}\epsilon^C_{~B}\rbrack/k^2.$$

\noindent Thus for the electromagnetic field for a $TM$ mode, Eq. (3.7), is 
$$\eqalignno {\lbrack F^k_{\mu \nu}\rbrack_{TM}&= \left[ \matrix {-\Phi \epsilon_{CB}Y^k
&-{\cal {A}}_BY^k_{,b}\cr
{\cal {A}}_BY^k_{,b}&0\cr}\right]&(3.12)\cr}$$
\vfill\eject

\centerline {\bf IV. ELECTROMAGNETIC FIELD DUE TO A PURELY LONGITUDINAL CURRENT.}

There is an equally good, if not slightly more direct way, of solving the vectorial $TM$ equations
(3.1).  Suppose the current is purely longitudinal, i.e.
$$\eqalignno {J^{\mu}&= (J^0, J^1, 0,0.)&(4.1)\cr}$$

\noindent This happens if, for example, a charge is accelerated linearly but otherwise quite 
arbitrarily.  This is the example in the next section where a uniformly accelerated Coulomb 
charge is considered.  For purely longitudinal currents such as these, the reduced current on
$M^2$ have according to Eq. (2.11) $$\eqalign {J&=0\cr
j&=0\cr}$$

\noindent but $j^A \not= 0$

\noindent The fact that $J=0$ implies that the $TE$ modes satisfy the homogeneous wave 
equation (3.2).  They don't interact with a longitudinal current.  The $TM$ modes on the other
hand satisfy the $TM$ wave Eqs. (3.1) with $j=0$.  With the help of Eqs. (3.8) and (3.9) they are
$$\eqalignno {- \lbrack {\cal {A}}_{E,F}\epsilon^{EF}\rbrack_{,C}\epsilon^C_{~B} + k^2  {\cal
{A}}_B &= 4\pi j_B&(4.2a)\cr {\cal {A}}^B_{\mid B} &= 0&(4.2b)\cr}$$

\noindent As a consequence, the charge current $j^B$ on $M^2$ also satisfies
$$\eqalignno {j^B_{\mid B}&= 0.&(4.3)\cr}$$

\noindent That ${\cal {A}}^B$ and $j^B$ have zero divergence implies (``Helmholz's theorem'') 
that there exist two respective scalars $\psi$ and $\eta$ on $M^2$ such that
$$\eqalignno {{\cal {A}}^B&= \psi_{,C}\epsilon^{CB}&(4.4a)\cr
j^B&= \eta_{,C}\epsilon^{CB}&(4.4b)\cr}$$

\noindent In terms of these scalars the $TM$ wave Eq. (4.2a) becomes with the help of Eq. (3.10)
$$\eqalignno {-\lbrack \psi_{,B}~^{\mid B}\rbrack_{,C} + k^2\psi_{,C}&= 4\pi j^B\epsilon_{BC}
&(4.5)\cr
&= 4\pi \eta_{,C}\cr}$$

\noindent Upon integration one obtains the $TM$ wave equation for a general longitudinal current,
Eq. (4.1a), 
$$\eqalignno {-\psi_{,B}~^{\mid B}+k^2\psi &= 4\pi \eta .&(4.5')\cr}$$

\noindent This $TM$ equation evidently has a structure identical to that of the $TE$ 
wave Eq. (3.2).  The scalar $TM$ equation yields the electromagnetic field mode, Eq. (3.12).  Its
form with the help of Eqs. (4.4a), (3.9), (4.5) and 
$${\cal {A}}_{E,F}\epsilon^{EF}= \psi^{\mid
D}_{,D}$$

\noindent is
$$\eqalignno {\lbrack F^k~_{\mu \nu}\rbrack_{TM}&= \left[ \matrix {-\psi^{\mid D}_{,D} 
\epsilon_{CB}Y^k&-\psi_{,C}\epsilon^C_{~B}~Y^k_{,b}\cr
~\psi_{,D}\epsilon^D_{~B}Y^k_{,b}&0\cr}\right]&(4.6)\cr}$$

\noindent This is a $TM$ electromagnetic field mode due to an arbitarily linearly 
moving charge distribution.  We shall now consider this $TM$ field due to a point charge 
in a linearly uniformly accelerating coordinate frame.

\vfill\eject

\centerline {\bf V. INTERACTION BETWEEN STATIC CHARGES.}

In particular, let us obtain in the accelerated frame what corresponds to the Coulomb field in an
inertial frame and thereby exhibit what corresponds to a multipole expansion in the inertial
frame.  This expansion in terms of the appropriate special functions is a natural consequence of the
$2+2$ decomposition of the Maxwell field.

The current $4$-vector of a point charge $e$ with four velocity $dz^\mu / ds$ is$^6$ 
$$\eqalignno {J^\mu (x^\nu )&= {e \over {\sqrt {-g}}}\int^\infty_{-\infty} {dz^\mu \over
{d\tau}}\delta^4\lbrack x^\nu - z^\nu (\tau )\rbrack d\tau&(5.1)\cr}$$

A charge which is static in a linearly uniformly accelerated frame has the world line
$$\eqalign {t&= \xi_0\sinh\,\tau\cr
x&=\xi_0 \cosh \,\tau\cr
y&=y_0\cr
z&= z_0\cr}$$

The current $4$-vector components relative to the coaccelerating basis are
$$\eqalignno {J^\tau &= {e \over {\xi}}\delta (\xi - \xi_0)\delta (y-y_0)\delta (z-z_0)&(5.2)\cr
J^\xi&= J^y = J^z = 0\cr}$$

It follows from Eq. (2.11) that the source for the reduced $TM$ and $TE$ wave equations is
$$\eqalignno {j^0&= {e \over {\xi}}\delta (\xi - \xi_0){e^{-i(k_yy_0+k_zz_0)} \over {2\pi
}}&(5.3a)\cr j^1&= 0&(5.3b)\cr
j&= J=0&(5.4)\cr}$$

Such a charge produces therefore no $TE$ electromagnetic field, i.e. $A=0$.  The charge 
current has no transverse components.  Consequently the $TM$ master Eq. (4.5) is applicable. 
Furthermore, the current and the field are time (``$\tau$'') independent.  Consequently the
$C=1$ component of the $TM$ Eq. (4.5) becomes with the help of Eq. (5.3a) and with $\epsilon_{01}
= \xi$ 
$$\eqalignno{-\left[{d^2\over d\xi^2}+{1\over \xi}{d\over d\xi}-\left({1\over
\xi^2}+k^2\right)\right] {d\psi \over {d\xi}} = 2e \delta (\xi - \xi_0)
e^{-i(k_yy_0+k_zz_0)}~.&&(5.5)\cr}$$


What are the boundary conditions that must be satisfied by the solution to this equation?  They
are determined by its physical significance.  The solution determines all non-zero (spectral)
components of the Maxwell tensor, Eq (4.6), the electric field
$$\eqalign{
[F^k_{10}]_{TM}=-{1\over \xi}{d\over d\xi}\xi{d\psi\over
d\xi}\epsilon_{01}Y^k(y,z)&\equiv -{\partial\over \partial \xi}\varphi_k(\xi)Y^k(y,z)\cr
[F^k_{y0}]_{TM}={d\psi\over d\xi}\epsilon_{10}Y^k_{,y}(y,z)&\equiv -{\partial\over \partial
y}\varphi_k(\xi)Y^k(y,z)\cr
[F^k_{z0}]_{TM}={d\psi\over d\xi}\epsilon_{10}Y^k_{,z}(y,z)&\equiv -{\partial\over \partial
z}\varphi_k(\xi)Y^k(y,z)\cr}$$

Thus 
$$\varphi_k(\xi)\equiv \xi{d\psi\over d\xi}$$
is the (spectral component of the) Coulomb potential, Eq. (4.4a),
$$\eqalign{{\cal A}_0&=-\varphi_k(\xi)\equiv -\xi{d\psi\over d\xi}\cr
{\cal A}_1&=0.\cr}$$
Relative to the physical (= orthonormal) basis $(\xi d\tau,d\xi)$ it satisfies the boundary
conditions
$$\eqalignno{&{d\psi\over d\xi}\to 0~~{\rm as}~~\xi\to\infty\quad\hskip1.1in \hbox{(spatial
infinity)}&(5.6)\cr &{d\psi\over d\xi}\to \hbox{  finite as  $\xi\to 0$\hskip 1in (event
horizon)}&(5.7)\cr}$$ It follows that the spectral potential obtained from the solution to Eq. (5.5)
is  
$$\eqalignno{\varphi_k(\xi)=\xi{d\psi\over d\xi}=2ec
e^{-i(k_yy_0+k_zz_0)}\,\xi\xi_0\,I_1(k\xi_{<})K_1(k\xi_{>})&&(5.8)\cr}$$

\noindent where 
$$I_1(k\xi_{<})K_1(k\xi_{>})\equiv \left\{ \matrix{I_1(k\xi)K_1(k\xi_0)&\xi<\xi_0\cr
I_1(k\xi_0)K_1(k\xi)&\xi>\xi_0.\cr}\right.$$
Here $K_1$ is the modified Bessel function which vanishes exponentially as $\xi\to\infty$, and
$I_1$ is the one which vanishes linearly as $\xi\to 0$.  The total Coulomb potential is obtained
as a sum from its spectral components, 
$$\eqalignno{\varphi(\xi,y,z)&=\int^{\infty}_{-\infty}\int^{\infty}_{-\infty}\varphi_k(\xi){e^{i(k_yy+k_zz)}\over
2\pi}dk_ydk_z&(5.9)\cr
&=2e\xi\xi_0\int^{\infty}_0I_1(k\xi_{<})K_1(k\xi_{>})\int^{2\pi}_0{e^{ikr\cos
(\theta-\alpha)}\over 2\pi}d\alpha\,kdk\cr }$$

Introduce the Bessel function 
$$J_0(kr)=\int^{2\pi}_0{e^{ikr\cos(\theta-\alpha)}\over 2\pi}d\alpha,$$
where 
$$\eqalign{&r=\sqrt{(y-y_0)^2+(z-z_0)^2},\quad r\cos \theta=y-y_0\cr
&k=\sqrt{k^2_y+k^2_z},\quad k\cos \alpha=ky.\cr}$$
Consequently, the potential becomes an integral involving the product of three Bessel functions 
$$\eqalignno{\varphi(\xi,y,z)=2e\xi\xi_0\int^{\infty}_0I_1(k\xi_{<})K_1(k\xi_{>})
J_0(kr)kdk.&&(5.10)\cr}$$

\noindent For $\xi<\xi_0$ the integral is$^7$
$$\int^{\infty}_0I_1(k\xi)K_1(k\xi_0)J_0(kr)kdk=(\xi\xi_0)^{-1}{-i\over
\sqrt{2\pi}}{Q^{1/2}_{1/2}(u)\over (u^2-1)^{1/4}}$$
where 
$$\eqalignno{Q^{1/2}_{1/2}(\cosh \gamma)&=i\left({\pi\over
2\sinh\gamma }\right)^{1/2}e^{-\gamma}&(5.11a)\cr \noalign{\hbox{and}}
u&={\xi^2+\xi^2_0+r^2\over 2\xi\xi_0}.&(5.11b)\cr}$$

Then the total potential is 
$$\eqalignno{\varphi(\xi,y,z)=e{u-(u^2-1)^{1/2}\over (u^2-1)^{1/2}}\qquad \xi<\xi_0&&(5.12)\cr}$$
For $\xi_0<\xi$ we expect the spectral integral to yield the same result.  Indeed, by resorting
to the identity$^8$ 
$$\int^{\infty}_0I_{\nu}(k\xi_0)K_{\nu}(k\xi)J_{\mu}(kr)k^{\mu+1}dk=e^{-i{\pi\over
2}\nu}\int^{\infty}_0I_{\nu}(ik\xi_0)J_{\nu}(k\xi)K_{\mu}(kr)k^{\mu+1}dk$$
one can use for $\xi_0<\xi$ the integral$^9$
$$\int^{\infty}_0I_1(ik\xi_0)J_1(k\xi)K_0(kr)kdk=(i\xi\xi_0)^{-1}{1\over
\sqrt{2\pi}}{Q^{1/2}_{1/2}(u)\over (u^2-1)^{1/4}}$$
to evaluate Eq. (5.10).  With the help of Eq. (5.11) the result is the same as Eq. (5.12), except
that $\xi_0<\xi$, as expected.

We conclude therefore that the Coulomb potential due to a static charge $e$ in a
(linearly uniformly) accelerated frame is 
$$\eqalignno{\varphi(\xi,\xi_0,y-y_0,z-z_0)&=e\int^{\infty}_{-\infty}\int^{\infty}_{-\infty}2\xi
\xi_0I_1(k\xi_{<})K_1(k\xi_{>})e^{i[k_y(y-y_0)+k_z(z-z_0)]}{d^2k\over 2\pi}\cr
&=e\left[{(\xi^2+R^2)^{{1\over 2}}\over R}-1\right]&(5.13)\cr}$$
where 
$$R^2=\xi^2(u^2-1)={[\xi^2_0+\xi^2+(x-x_0)^2+(y-y_0)^2]^2-4\xi^2\xi^2_0\over 4\xi^2_0}.$$

\noindent We can express the potential (5.13) in terms of the complete, orthonormal set of Bessel
harmonics
$$\left\{J_m(kr){e^{im\varphi}\over \sqrt{2\pi}}\right\}.$$
In terms of these, the plane wave harmonics are 
$${e^{i[k_y(y-y_0)+k_z(z-z_0)]}\over 2\pi}=\sum^{\infty}_{m=-\infty}i^m{e^{im(\theta-\alpha)}\over
2\pi}J_m(kr).$$
Introducing this into (5.13) we recover Eq. (5.10), 
$$\varphi(\xi, \xi_0,y-y_0,z-z_0)=e\int^{\infty}_02\xi\xi_0I_1(k\xi_{<})K_1(k\xi_{>})J_0(kr)k\,dk.$$
This representation is analogous to the familiar multipole expansion,
$${e\over\sqrt{(x-x_0)^2+(y-y_0)^2+(z-z_0)^2}}=e\sum^{\infty}_{\ell=0}r_{<}^{\ell}~
r_{>}^{-\ell-1}P_{\ell}(\cos\theta),$$
of an inertial charge.  The ``multipoles'' (if one insists on introducing them) for the charge
static in the accelerated frame are evidently characterized by the continuous index $k$ instead of
the discrete index $\ell$ for the inertial case.

\vfil\eject

\centerline{\bf VI. SPECTRAL REPRESENTATION.}

Quantum mechanically the Coulomb energy between a pair of static charges is a spectral sum of
processes in each one of which a pair of virtual quanta is exchanged.  

In order to motivate the extension of this spectral decomposition from an inertial frame to a
uniformly accelerated frame, let us recall its quantum mechanical basis relative to the inertial
frame.  The key features are already contained in the simpler Yakawa interaction, which is mediated
by a scalar field instead of the vectorial Maxwell field.

Between two heavy inertial non-relativistic nucleons situated at $\vec X_1$ and $\vec X_2$ the
interaction is 
$$\eqalignno{\Delta E(\vec X_1-\vec X_2)=-{g^2\over 4\pi}{e^{-m\vert\vec X_1-\vec X_2\vert}\over
\vert\vec X_1-\vec X_2\vert}&&(6.1)\cr}$$
It is the quanta of the scalar field $\phi$, 
$$(\bx+m^2)\phi=0,$$
which mediates this interaction.  Indeed, the Hamiltonian for the interaction between this meson
field and the two heavy mucleons is 
$$\eqalignno{H_{int}=g\int[F(\vec X_1-\vec Y)+F(\vec X_2-\vec Y)]\phi(\vec
Y)d^3y.&&(6.3)\cr}$$

\noindent Here $g$ is the coupling constant (``charge'') of a nucleon, and $F$ expresses the
finiteness of a nucleon, which in the limit of a point charge yields
$$F(\vec X_1-y)\to \delta^3(\vec X,-Y).$$
The interaction Hamiltonian perturbs the lowest energy state of the unperturbed Hamiltonian
$$\eqalignno{H_0=\int \omega(\vec k)a^{\ast}_{\vec k}a_{\vec k}d^3k+2M&&(6.4)\cr}$$
of the system:  2 nucleons each of rest mass $M$ together with the meson field whose quanta have
energy $\omega(\vec k)$.  The perturbation in the lowest energy state is determined by second order
perturbation theory, and it is given by 
$$\eqalignno{\Delta E=\sum_n{\langle 2,0\vert H_{int}\vert n\rangle\langle n\vert H_{int}\vert
2,0\rangle\over 2M-E_n}.&&(6.5)\cr}$$

\noindent The meson field operator
$$\eqalignno{\phi={1\over (2\pi)^{3/2}}\int\!\!\int\!\!\int {a_ke^{-i\omega_kt}\over
\sqrt{2\omega_k}}e^{ik\cdot x}d^3k+~{\rm herm. adj.}&&(6.6)\cr}$$
in the interaction Hamiltonian implies that the only intermediate states contributing to $\Delta E$
are those consisting of two nucleons plus one meson.  Making use of 
$$\eqalignno{\sum_n{(\cdots)\vert n\rangle \langle n\vert  (\cdots)\over 2M-E_n}\to
\int\!\!\int\!\!\int {(\cdots)\vert 2,1_k\rangle\langle 2,1_k\vert(\cdots)\over
-\omega_k}d^3k,&&(6.7)\cr}$$
a modest amount of algebra yields
$$\eqalignno{\Delta E=-{g^2\over (2\pi)^3}\int\!\!\int\!\!\int \vert\widetilde F(k)\vert^2{1\over
2\omega_k}\{1+1+e^{ik\cdot (x_1-x_2)}+e^{ik(x_2-x_1)}\}d^3k.&&(6.8)\cr}$$

\noindent Thus the perturbation $\Delta E$ arises from four processes involving the emission and
reabsorption of quanta.  See Figure 1a and 1b.  In the first, nucleon number one emits and absorbs a
quantum.  In the second, nucleon number two does the same.  In the third and fourth, a quantum which
is emitted by one nucleon gets absorberd by the other.  Thus the perturbation $\Delta E$ decomposes
into 
$$\eqalignno{\Delta E=\Delta E_1+\Delta E_2.&&(6.9)\cr}$$
Here the exchange energy is 
$$\eqalignno{\Delta E_2&=-{g^2\over (2\pi)^3}\int\!\!\int\!\!\int {e^{ik\cdot(X_1-X_2)}\over
k^2+m^2}d^3k&{\rm (6.10a)}\cr
&=-{g^2\over 4\pi}{e^{-m\vert X_1-X_2\vert}\over \vert X_1-X_2\vert}&{\rm (6.10b)}\cr}$$
One can probably give an analogous succinct derivation for an accelerated frame, but in that case
additional qualitative features enter.  See Section VIII.  The purpose of this present section is to
give in the accelerated frame a spectral decomposition of the Coulomb energy
$$\eqalignno{V(\xi,\xi_0,y-y_0,z-z_0)=e^2\left[{\xi^2+R^2\over R}-1\right],&&(6.11)\cr}$$
a decomposition which parallels the one for an inertial frame, Eq. (6.10).

A normal mode solution to the wave equation in an accelerated frame is 
$$e^{-i\omega\tau}K_{i\omega}(k\xi){e^{i(k_yy+k_zz)}\over 2\pi}$$
and the corresponding wave is 
$$\eqalignno{K_{i\omega}(k\xi){e^{i(k_yy+k_zz)}\over 2\pi}.&&(6.12)\cr}$$
Like their inertial cousins in Eq. (6.10a), these waves are orthonormal and they form a complete set.

Indeed, the longitudinal ($\xi$-dependent) part of this wave satisfies the orthogonality
relation,
$$\eqalignno{\int^{\infty}_0{2\omega\over
\pi^2}\sinh\pi\omega K_{i\omega'}(k\xi)K_{i\omega'}(k\xi){d\xi\over \xi}=[\delta
(\omega-\omega')+\delta(\omega+\omega')],&&(6.13)\cr}$$
the completeness relation,
$$\eqalignno{\int^{\infty}_0{1\over
\xi}K_{i\omega}(k\xi)K_{i\omega}(k\xi'){\omega\sinh\pi\omega\over
\pi^2}d\omega=\delta(\xi-\xi'),&&(6.14)\cr}$$

\noindent ($K_{i\omega}$ is an even function of $\omega$) and the differential equation 
$$\eqalignno{-\left[{d\over d\xi}\xi{d\over d\xi}-k^2\xi\right]K_{i\omega}(k\xi)=
{\omega^2\over \xi}K_{i\omega}(k\xi).&&(6.15)\cr}$$
Applying the completeness relation to the right hand side of Eq. (5.5), expanding the solution to
Eq. (5.5) in terms of the longitudinal wave function $K_{i\omega}(k\xi)$, using Eq. (6.15), 
and finally using the orthogonality relation Eq. (6.13), one obtains 
$$\eqalignno{\varphi_k(\xi)\equiv\xi{d\psi\over
d\xi}=2ee^{-i(k_yy_0+k_zz_0)}\int^{\infty}_0{2\omega\sinh\pi\omega\over \pi^2}{\xi
K_{i\omega}(k\xi)K_{i\omega}(k\xi_0)d\omega\over \omega^2+1}&&(6.16)\cr}$$
This result can also be obtained without using the completeness of the set of wave functions
$K_{i\omega}(k\xi)$.  One simply accepts a well-documented integral$^{10}$ to replace the
product in Eq. (5.8) with its spectral representation, Eq. (6.16).

The total Coulomb potential is now given by Eq. (5.13).  The spectral decomposition of the
Coulomb energy $V=e\varphi$ between two static charges in a linearly accelerated frame is
therefore
$$\eqalignno{&V(\xi,\xi_0,y-y_0,z-z_0)=e^2\left[{(\xi^2+R^2)^{1/2}\over R}-1\right]\cr
&=2e^2\int^{\infty}_0\int^{\infty}_{-\infty}\int^{\infty}_{-\infty}{2\omega\sinh\pi\omega\over\pi^2}{\xi
K_{i\omega}(k\xi)K_{i\omega}(k\xi_0)\over \omega^2+1}{e^{i[k_y(y-y_0)+k_z(z-z_0)]}\over
2\pi}d\omega\,dk_y\,dk_z.&(6.17)\cr}$$

This is the interaction which is analogous to the Coulomb interaction, Eq. (6.10), in an
inertial frame.
 \vfil\eject

\centerline{\bf VII.  RADIATION REACTION}

With a fixed (at $\xi_0, y_0,z_0$) charge $e$ giving rise to its static Coulomb potential 
$$\eqalignno{\varphi(\xi,\xi_0,y-y_0,z-z_0)=e\left[{\xi^2_0+\xi^2+r^2\over
\sqrt{(\xi_0^2+\xi^2+r^2)^2-4\xi^2\xi^2_0}}-1\right]&&(7.1)\cr}$$
one might wonder:  Where is the emitted Larmor radiation?  We shall answer this question with a 
heuristic argument which is based on the idea that the future event horizon is the history of a
two-dimensional resistive membrane$^{11-13}$.

Recall the Lorentz-Dirac equation of motion of a point particle of mass $m$ and charge $e$ under the
action of an external force $F^{\mu}$:
$$\eqalignno{{d\over ds}\left( m{dx^{\mu}\over ds}-{2\over 3}e^2{d^2x^{\mu}\over ds^2}\right)=-{2\over
3}e^2{dx^{\mu}\over ds}{d^2x^{\alpha}\over ds^2}{d^2x^{\beta}\over
ds^2}g_{\alpha\beta}+F^{\mu}.&&(7.2)\cr}$$
This equation together with its physical interpretation$^{14,15-17}$ is a direct consequence of the
conservation of total momentum-energy (=``momenergy''$^{18}$), electromagnetic together with
mechanical, of the particle along its worldline.  

Recall that the electromagnetic momenergy of a charge splits unambiguously into two mutually
exclusive and jointly exhaustive parts:~~(1)~~that which is {\it radiated} and  (2) that which is
{\it bound}$^{14}$ to the charge.  Each of these two parts is determined by its own respective
electromagnetic stress-energy tensor, both of which are divergenceless everywhere except on the
worldline of the charge.  The identifying feature of the radiated stress-energy tensor is that (a) it
is quadratic in the acceleration and that (b) its form is that of a simple null fluid, even close to
the charge$^{14}$.

The purely mechanical momenergy describes the ``bare'' (i.e. without any electromagnetic field)
particle.  Its stress-energy tensor is also divergenceless everywhere except on the particle
worldline.

As for any total stress-energy tensor, the sum of its three individual sources vanishes.  In fact,
the sources and sinks of the three stress-energies are balanced perfectly along the whole history of
the particle.  This balance is expressed by the Lorentz-Dirac Eq. (7.2).

Its left hand side is the rate of change of momenergy
$$\eqalignno{P^{\mu}=m{dx^{\mu}\over ds}-{2\over 3}e^2{d^2x^{\mu}\over ds^2}.&&(7.3)\cr}$$
It is the sum of the particle's inertial momenergy $(\propto {dx^{\mu}\over ds})$ and the
electromagnetic momentum-energy $(\propto {d^2x^{\mu}\over ds^2})$ which is always bound to the
charge no matter what its instantaneous state of acceleration might be.  Although one often talks 
about these two momenergies separately, physically the two go together.

The right hand side, apart from $F^{\mu}$, is the Larmor expression for the rate at which the
charged particle loses momenergy in the form of e.m. radiation.  The Lorentz-Dirac equation demands
that this radiation reation four-force change the inertial plus bound e.m. momenergy of the charged
particle.  The magnitude of the radiation four-force is given by the invariant Larmor formula,
$$\eqalignno{{\rm power~}={2\over 3}e^2{d^2x^{\alpha}\over ds^2}{d^2x^{\beta}\over
ds^2}g_{\alpha\beta}.&&(7.4)\cr}$$

Let us illustrate an alternative viewpoint by deriving this power for a Coulomb charge static in a
linearly uniformly accelerated coordinate frame.  The line of reasoning goes roughly like this:  Let
the charge be fixed at $(\xi_0,y_0,z_0)$ in the accelerated frame.  The charge is surrounded by
equipotential surfaces, Eq. (7.1), $$\varphi(\xi,\xi_0,y-y_0,z-z_0)=e\left[{(\xi^2+R^2)^{1/2}\over
R}-1\right]={\rm ~constant}.$$ The electric field
$$\eqalignno{E_{\xi}&\equiv F_{\xi\mu}u^{\mu}=\vert g^{\tau\tau}\vert^{1/2}F_{\xi 0}={-1\over
\xi}{d\varphi\over d\xi}=-e{4\xi^2_0(\xi^2_0+r^2-\xi^2)\over
[(\xi^2+\xi^2_0+r^2)^2-4\xi\xi_0]^{3/2}}\cr
E_y&\equiv F_{y\mu }u^{\mu}=\vert g^{\tau\tau}\vert^{1/2}F_{y0}={-1\over \xi}{d\varphi\over
dy}=e{\xi\over \xi_0}{(y-y_0)\cdot 8\xi^3_0\over [(\xi^2+\xi^2_0+r^2)^2-4\xi^2_0\xi^2]^{3/2}}\cr
E_z&\equiv F_{z\mu }u^{\mu}=\vert g^{\tau\tau}\vert^{1/2}F_{z0}={-1\over \xi}{d\varphi\over
dz}=e{\xi\over \xi_0}{(z-z_0)\cdot 8\xi^3_0\over
[(\xi^2+\xi^2_0+r^2)^4-4\xi^2_0\xi^2]^{3/2}}&(7.5)\cr}$$
is perpendicular to these potential surfaces and hence also to the event horizon $\xi=0$ where
$\varphi=0$.  There this electric field induces the surface charge density$^{19,20,21}$
$$\eqalignno{\sigma={1\over 4\pi}E_{\xi}\vert_{\xi=0}=-{e\over \pi}{\xi^2_0\over
(\xi^2_0+r^2)^2}.&&(7.6)\cr}$$
The total charge induced at the event horizon is 
$$\eqalignno{\int\limits^{\infty}_{-\infty}\!\!\!\int\limits^{\infty}_{-\infty}\sigma
dy\,dz=-e.&&(7.7)\cr}$$
This, by the way, supports the fact the event horizon behaves like the history of a conductive
surface$^{11-13}$.

Electrostatics implies that there is an attractive force between the point charge $e$ and the
surface charge density $\sigma$.  From symmetry this force is directed along the $\xi$ direction and
has magnitude
$$\eqalignno{\vert{\rm Force}\vert&={1\over 2}\int\!\!\!\int E_{\xi}\sigma\vert_{\xi=0}dy\,dz\cr
&=\int\!\!\!\int{E^2_{\xi}\over 8\pi}\vert_{\xi=0}dy\,dz\cr
&={2\over 3}{e^2\over \xi^2_0}\cr
\noalign{{\rm or}}&\cr
{\rm power }&={2\over 3}e^2\,{\rm (acceleration)}^2&(7.8)\cr}$$
in terms of relativistic units, which we are using.  This force, or power, is a rate of change;
furthermore, it is a rate of change which is tangential to the future event horizon.  It therefore
expresses a flow of momenergy across the two dimensional membrane (spanned by $y$ and $z$)
whose history is that future event horizon.  Conservation of momenergy (``for every action
there is an opposite and equal reaction'') implies that the momenergy gets drained from the
charged particle at a rate, which yields the magnitude of the radiation reaction on the right hand
side of the Lorentz-Dirac Eq. (7.2).  

Identifying the momenergy in the event horizon with radiation momenergy losses along the particle
worldline hinges on a tacit assumption:  None of the particle's {\it bound}$^{14}$ stress-energy
tensor enters into the momenergy conservation between the particle's worldline and its future event
horizon.

The fact that the rate given by Eq. (7.8) agrees with Larmor's formula (7.4) leads us into making
two observations.

\noindent 1.~~~The existence of charge density on the event horizon is a purely observer dependent
phenomenon.  Relative to an inertial observer there is, of course, no charge on the event horizon of
the accelerated corrdinate frame.  On the other hand, relative to the accelerated frame, the nature
of the interaction of matter with radiation manifests itself such that there is no way of
distinguishing the field of this charge density from that of an actual charge.

\noindent 2.~~~If we extend our considerations of a Coulomb field static in an accelerated
frame to those of a moving charge, then the possibility exists of having surface currents as well as
intrinsic electric and magnetic fields evolve on the event horizon.  They give rise, among others,
to resistive forces$^{11}$ which act back on their sources and thus could give a picturesque account
of the radiation reaction force of a point charge.  Such a picturesque account can often be given by
referring to a black hole analogue.  An obvious example is a charged particle suspended above the
equator of a rotating black hole.  This can be viewed as the electrostatic analogue of a black hole
rotating in an oblique magnetic field$^{22}$.  A resistive spin-down torque is exerted on the black
hole$^{11}$.  The back reaction on the charge can be viewed as the radiation reaction which enters
the Lorentz-Dirac Eq. (7.2).  This one can do, provided one replaces the space-time of the rotating
black hole with the appropriate flat space-time analogue:  The coordinate frame of a linearly
accelerated observer with uniform transverse drift$^{23}$.  

Although one can believe that the back
reaction from the black hole corresponds to the radiation reaction in drifting Rindler spacetime, the
analogue of the increase of the entropy of a black hole$^{24}$ is still rather murky.  This is so
because there does not yet exist for Rindler spacetime a definition of what in gravitation physics is
called a black hole entropy$^{24}$.
 
That such an analogue should exist is not entirely unreasonable if one recalls that radiation losses
from an accelerated charge are resistive and hence irreversible in nature.  It presumably is this
irreversibility which would be expressed by an increase in the to-be-defined entropy of a Rindler
event horizon.

\vfil\eject

\centerline{\bf VIII.  ACCELERATED POLARONS?}

Compare the Coulomb interaction energy between a pair of charges static in an accelerated frame
as given by Eq. (6.17), with the Yukawa interaction energy between a pair of charges static in
an inertial frame,
$$\eqalignno{V(\vec X_1-X_2)&={e^2\over 4\pi}{e^{-m\vert\vec X_1-\vec X_2\vert}\over
\vert \vec X_1-\vec X_2\vert}\cr
&=e^2\int\!\!\int\!\!\int {1\over k^2_x+k^2_y+k^2_z+m^2}e^{i(k_xx+k_yy+k_zz)}{dk_xdk_ydk_z\over
(2\pi)^3}.&(8.1a)\cr}$$ 
This interaction energy is attributed to a second order process which expresses the
exchange of quanta of momentum $\hbar (k_x,k_y,k_z)$ between two non-relativistic
charges, each of rest mass $m$.  The emission and absorption of these virtual ``Minkowski'' quanta
takes place in an inertial frame where the ground state of the field is the familiar Minkowski vacuum.

In an accelerated frame, however, the elementary excitations are not the Minkowski quanta. 
Instead one has the Fulling quanta$^{25}$.  They are elementary excitations of a different
ground state, the Rindler vacuum.  This ground state consist of a configuration of highly
correlated photons$^3$.

The problem of the Coulomb interaction between a pair of charges accelerating through this
correlated photon configuration is analogous to the interaction between a pair of charges in a
polar crystal (``interaction between a pair of polarons'').  In such a crystal a single charge
is referred to as a ``polaron'' because it consists of the charge together with the local
lattice polarization distortion which the charge produces$^{26}$.  This affects its mass (``mass
renormalization'') and its interaction with other charges$^1$ (``coupling constant
renormalization'').  An accelerated charge in a correlated photon configuration may be viewed in
the same way.  The charge distorts this correlated configuration (the ``Rindler vacuum state'')
and gives rise to an ``accelerated polaron''.  Consider the interaction between two such
``accelerated polarons''.   The description of this interaction in terms of photons is
complicated when it is done in terms of photons, just as the interaction between two polarons
in a crystal is complicated when done in terms of the lattice atoms.  A much more natural
description is in terms of elementary excitations.  For ``accelerated polarons'' this means a
description in terms of (virtual) Fulling quanta, just like for crystal polarons this means a
description in terms of (virtual) sound quanta.  These quanta have however an effect on the Coulomb
interaction$^1$.  Instead of Eq. (8.1), the interaction potential (with $m=0$) is 
$$V(\vec X_1-\vec X_2)={1\over\epsilon}{e^2\over 4\pi}{1\over \vert\vec X_1-\vec X_2\vert}.$$
The potential is still of the Coulomb type, but with a change in the coupling constant 
$$e^2\to e^2/\epsilon$$
due to the dielectric constant $\epsilon$ of the crystal.

This analogy with polarons in a crystal suggests an inquiry as to whether the Coulomb interaction
between two accelerating charges is also altered.  In other words, does the classical Coulomb
potential, Eq. (5.13) or (6.17), differ from the Coulomb potential determined quantum mechanically by
a factor which expresses the ``polarizibility'' and hence the dielectric constant of the Rindler
vacuum?
\bigskip

\noindent{\bf Acknowledgements}
\medskip

F. J. A. was supported in part by a Senior Honors Research Scholarship at the Ohio State
University by a Rutgers University Excellence Fellowship, and at Rutgers by NSF Grant DMR
89-1893.  F. J. A. thanks Joel Lebowitz for his support during the completion of this project.
\bigskip

\centerline{\bf References}
\medskip

\ref{[1]}{See F. Strocchi, {\it Elements of Quantum Mechanics of Infinite Systems} (World
Scientific, Singapore, 1985), Ch. II, for an intuitive picture.}

\ref{[2]}{S. Schweber, Relativistic Quantum Field Theory (Harper and Row) Ch. 12.}

\ref{[3]}{U. H. Gerlach, {\it Phys. Rev. D.} {\bf 40} 1037 (1989).}

\ref{[4]}{U H. Gerlach, Ann. Inst. Henri Poincar\'e, \underbar{49}, 397 (1988); Ann. N.Y. Acad. 
Sci. \underbar{571}, 331 (1989); in {\it Proceeding of the 5th Marcel Grossmann Meeting} edited by D.
G. Blair and M. J. Buckingham (World Scientific Publishing Co., New Jersey, 1989) p. 797.}

\ref{[5]}{Except for a scale factor, these $TM$ equation are the same as the $(-1)^{\ell+1}$ parity
linearized Einstein field equations on an arbitrary spherically symmetric background.  See U. H.
Gerlach and J. F. Scott Phys. Rev. D 34, 3638 (1986).  The scale factor, $r^2$, is introduced by
making the substitutions $k^2\to (\ell-1)(\ell+2)/r^2$, ${\cal A}_{C\vert B}\,^{\vert C}\to
r^{-2}[r^4(r^{-2}k_C)_{\vert B}]^{\vert C}$ and similarly for $A_{B\vert C}\,^{\vert C}$.  Evidently
for large radial coordinate $r$, an $(-1)^{\ell+1}$ parity gravitational disturbance is the same as an
electromagnetic $TM$ mode.}

\ref{[6]}{J. D. Jackson, {\it Classical Electrodynamics} (John Wiley and Sons, New York, 1975).  Sec.
14.1.}

\ref{[7]}{I. S. Gradshteyn and I. M. Ryzhik, {\it Table of Integrals Series and Products} (Academic 
Press, New York, 1980), p. 696, Eq. 11.}

\ref{[8]}{Bailey, Proc. Lond. \underbar{40}, 43 (1936).}

\ref{[9]}{I. S. Gradshtein and I. M. Ryzhik, ibid, p. 695, Eq. 7.}

\ref{[10]}{See Eq. (10) in I. S. Gradstein and I. M. Ryzhid, ibid, p. 773.}

\ref{[11]}{T. Damour, Phys, Rev. D \underbar{18}, 3598 (1978).}

\ref{[12]}{R. L. Znajek, Mon. Not. R.A.S. \underbar{185}, 833 (1978).}

\ref{[13]}{T. Damour in {\it Proceedings of the Second Marcel Grossmann Meeting on General
Relativity (Trieste, 1979)}, Part A, edited by R. Ruffini (North Holland, New York, 1982) p.
587-608.}

\ref{[14]}{C. Teitelboim, Phys. Rev. D \underbar{1}, 1572 (1970).  This is a very useful paper. 
Among other things, it introduces the concept of electromagnetic $4$-momentum ``bound'' to a
charge.  It also delineates the key ideas of the pioneering works of P.A.M. Dirac and F.
Rohrlich on the subject matter.}

\ref{[15]}{C. Teitelboim, Phys. Rev. D \underbar{3}, 297 (1971).}

\ref{[16]}{C. Teitelboim, Phys. Rev. D \underbar{4}, 345 (1971).}

\ref{[17]}{C. Teitelboim, D. Villarroel, and Ch. G. van Weert, Rivista Del Nuovo Cimento
\underbar{3}, 1 (1980).}

\ref{[18]}{The contraction of ``momentum-energy'' into ``momenergy'' is rather natural.  For a
rationale see J. A. Wheeler, {\it A Journey into Gravity and Spacetime} (W. H. Freeman and Co., New
York, 1990), Ch. 6.}

\ref{[19]}{The concept of charge density induced on a black hole was first introduced by R. S. Hanni
and R. Ruffini, Phys. Rev. D 8, 3259 (1972).}

\ref{[20]}{D. A. Macdonald and W.-Mo Suen, Phys. Rev. D \underbar{32}, 848 (1985) give the electric
and magnetic fields of a charge in various states of motion in an accelerated frame.}

\ref{[21]}{Ch. 2 in {\it Black Holes:  The Membrane Paradigm} by K. S. Thorne, R. H. Price, and D.
A. Macdonald (Yale University Press, New Haven, 1986) gives an introductory treatment of the
electrical properties of the surface whose history is an event horizon.}

\ref{[22]}{W. H. Press, Astrophys. J. \underbar{175}, 243 (1972).}

\ref{[23]}{U. H. Gerlach, Phys. Rev. D \underbar{27}, 2310 (1983).}

\ref{[24]}{J. D. Bekenstein, Phys. Rev. D \underbar{7}, 2333 (1973).}

\ref{[25]}{S. A. Fulling, Phys. Rev. D\underbar{7}, 2850 (1973).}

\ref{[26]}{R. P. Feynman, {\it Statistical Mechanics:  A Set of Lectures} (Benjamin/Cummings
Publishing Co., Reading, Mass. 1982), Ch. 8.}

\end

\ref{[11]}{D. G. Boulware, Ann. Physics (New York), {\bf 124}, 169 (1980).}

\ref{[12]}{S. A. Fulling, Phys. Rev D\underbar{7}, 2850 (1973).}

\ref{[13]}{R. P. Feynman, {\it Statistical Mechanics:  A Set of Lectures} (Benjamin/Cummings
Publishing Co., Reading, Mass. 1982), Ch. 8.}

\end

\centerline {\bf ABSTRACT}
\bigskip

The inhomogeneous Maxwell wave equations are adopted to a coordinate frame which is accelerating
linearly but arbitrarily.  The symmetry along the transverse direction permits a reduction of the
variational principle and its concomitant field equations to the $2$-$D$ Lorentz spacetime of the
accelerated frame.  The equations for the potentials are gauge invariant and separate on this $2$-$D$
Lorentz manifold into a vector equation which governs the $TM$ modes, and a scalar equation which
governs the $TE$ modes. 

We solve these equation for a static charge in a uniformly accelerated frame.  A spectral
decomposition of the Coulomb potential is obtained and then integrated to recover the Coulomb
potential in the $4$-$D$ space-time of the accelerated coordinates.  A comparison is drawn with the
Coulomb interaction between a pair of polarons in a crystal and contrasted with the familiar
Coulomb potential between two inertial charges.


\noindent This is the $(k_y, k_z)$ Fourier component of the Poisson equation
$$\lbrace {1 \over {\xi}} {\partial \over {\partial \xi}}{1 \over {\xi}} {\partial \over 
{\partial \xi}} + {\partial^2 \over {\partial y^2}} + {\partial^2 \over {\partial z^2}}\rbrace
V(\xi , y, z) = -4\pi ec {\delta (\xi - \xi_0) \over {\xi}} \delta (y-y_0) \delta (z-z_0)$$

For the total Colomb potential 
$$\eqalign {V(\xi , y, z)&= \int \int {\cal {A}}^{0k} \epsilon_{01}Y^k(y,z)d^2k\cr
&= \xi \int \int {d\psi \over {d\xi}}~~Y^k (y,z)d^2k\cr}$$

\noindent in the accelerated coordinate frame.

From this potential one obtains with the help of Eq. (4.6) all non-zero components of the 
electromagnetic tensor
$$\eqalign {F_{01}&= - {\partial \over {\partial \xi}}V(\xi , y, z)\cr
F_{0y}&= - {\partial \over {\partial y}} V(\xi , y,z)\cr
F_{0z} &= - {\partial \over {\partial z}} V(\xi , y, z).\cr}$$

The interaction energy between the charge $e$ located at $(\xi_0, y_0, z_0)$ and a second 
charge $q$ located at $(\xi , y, z)$ is
$$E = q V(\xi , y, z)$$